\pdfoutput=1
\documentclass[11pt]{article}

\usepackage[preprint]{acl}

\usepackage{amsmath}
\usepackage{enumitem}
\usepackage{multirow}
\usepackage{booktabs}
\usepackage{array}
\usepackage{diagbox}
\usepackage{colortbl}
\PassOptionsToPackage{dvipsnames}{xcolor}
\usepackage{verbatim}

\usepackage{times}
\usepackage{latexsym}

\usepackage[T1]{fontenc}
\usepackage[utf8]{inputenc}

\usepackage{microtype}

\usepackage{inconsolata}

\usepackage{graphicx}

\title{PodAgent: A Comprehensive Framework for Podcast Generation}
\author{
  \textbf{Yujia Xiao\textsuperscript{1}},
  \textbf{Lei He\textsuperscript{2}},
  \textbf{Haohan Guo\textsuperscript{1}},
  \textbf{Fenglong Xie\textsuperscript{3}},
  \textbf{Tan Lee\textsuperscript{1}}
\\
\\
  \textsuperscript{1}The Chinese University of Hong Kong,
  \textsuperscript{2}Microsoft,
  \textsuperscript{3}Xiaohongshu Inc.
\\
  \small{
    \{\href{mailto:yujiaxiao@link.cuhk.edu.hk}{yujiaxiao@link.cuhk.edu.hk},
    \href{mailto:helei@microsoft.com}{helei@microsoft.com}\}
  }
\\
  \small{
    \{\href{mailto:hguo@se.cuhk.edu.hk}{hguo@se.cuhk.edu.hk},
    \href{mailto:fenglongxie@xiaohongshu.com}{fenglongxie@xiaohongshu.com},
    \href{mailto:tanlee@ee.cuhk.edu.hk}{tanlee@ee.cuhk.edu.hk}\}
  }
}

\def\model{PodAgent}

\begin{document}
\maketitle
\begin{abstract}
Existing automatic audio generation methods struggle to generate podcast-like audio programs effectively. The key challenges lie in in-depth content generation, appropriate and expressive voice production. This paper proposed \model, a comprehensive framework for creating audio programs. \model\ 1) generates informative topic-discussion content by designing a Host-Guest-Writer multi-agent collaboration system, 2) builds a voice pool for suitable voice-role matching and 3) utilizes LLM-enhanced speech synthesis method to generate expressive conversational speech. Given the absence of standardized evaluation criteria for podcast-like audio generation, we developed comprehensive assessment guidelines to effectively evaluate the model's performance. Experimental results demonstrate \model's effectiveness, significantly surpassing direct GPT-4 generation in topic-discussion dialogue content, achieving an 87.4\% voice-matching accuracy, and producing more expressive speech through LLM-guided synthesis. Demo page: \href{https://podcast-agent.github.io/demo/}{https://podcast-agent.github.io/demo/}. Source code: \href{https://github.com/yujxx/PodAgent}{https://github.com/yujxx/PodAgent}.

%Experimental results indicate that the topic-discussion dialogue content produced by \model significantly outperforms that generated directly by GPT-4. Furthermore, the voice-matching rate reached 86.5\%, and the generated speech—optimized by LLM-predicted instructions—also achieved better expressiveness.

\end{abstract}

\section{Introduction}

\begin{figure}[t]
  \includegraphics[width=\columnwidth]{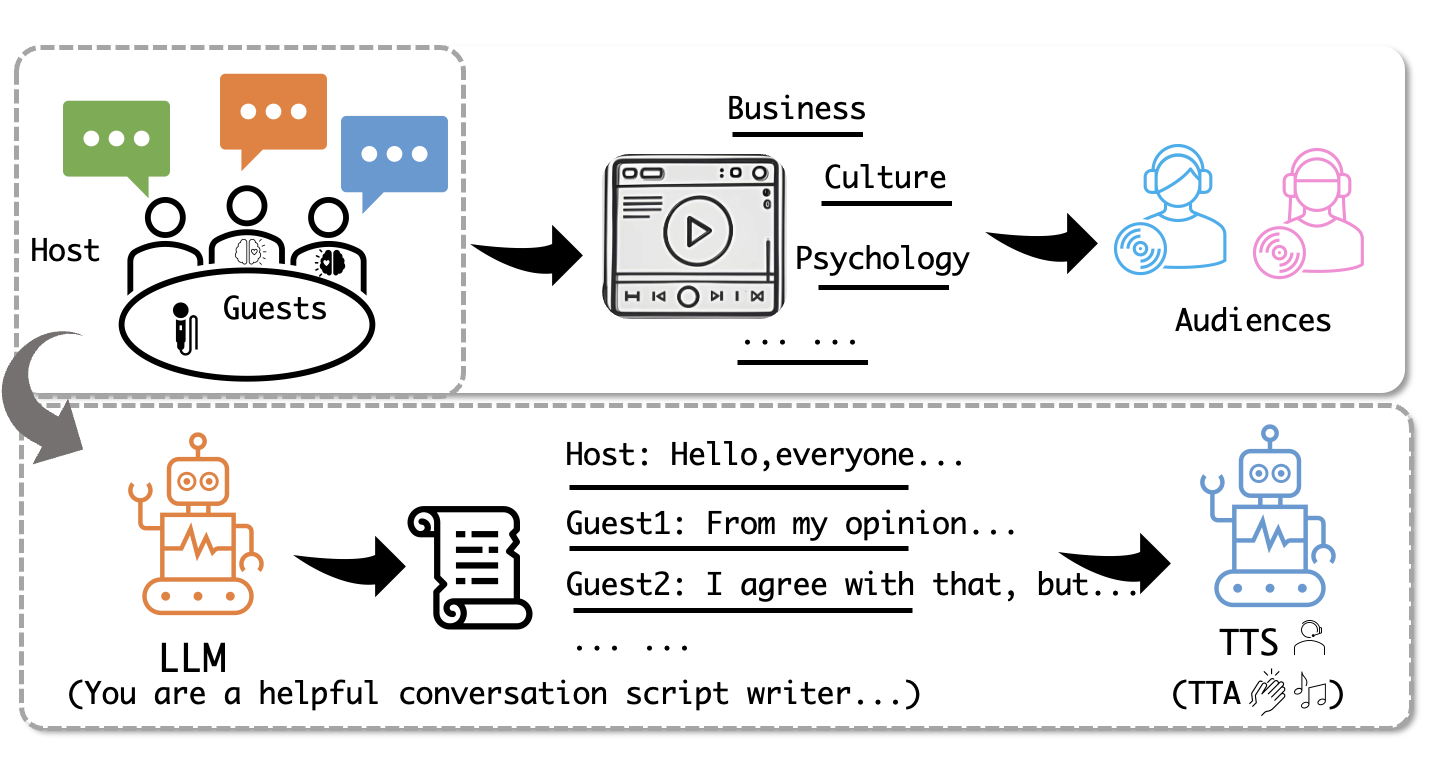}
  \caption{An overview of human-made Podcasts (Above) and the Generative models based \model\ (Below): LLMs / TTS / TTA are used to generate conversation scripts, speech, sound effect and music. }

  \label{fig:concept}
\end{figure}

Audio programs are an important channel for information acquisition. Compared to video or text media, audio can free your eyes and hands, allowing you to access information conveniently in a variety of scenarios. Podcasts are digital audio programs that are available for streaming or downloading over the Internet. As shown in Figure ~\ref{fig:concept} (Above), the content of a podcast typically consists of viewpoints shared by various individuals. To accommodate diverse interests, podcasts often explore a wide range of topics, including economics, culture, psychology, and more. However, many content creators still face the complex process of transforming creative ideas into a final product. Additionally, providing strong, well-founded viewpoints and producing high-quality podcast-like audio on any given topic remains a significant challenge. 

Recent advancements in generative models have made it possible to automatically create high-quality content. Large language models (LLMs) \cite{ouyang2022,achiam2023gpt,team2023gemini,touvron2023llamaopenefficientfoundation, anthropic_claude} have achieved breakthrough capabilities in generating coherent and contextually appropriate text. In addition, foundation models for other modalities, such as vision \cite{blattmann2023stable, midjourney2023,brooks2024video} and audio \cite{borsos2023audiolm,wang2023neural,bark2023, copet2024simple}, have made remarkable strides in the creation of multimodal content.

While existing models can generate podcast-like content, they have not yet achieved the creation of complete, professionally structured podcast episodes. For instance, audio-enhanced multimodal LLMs \cite{wu2023next, huang2023audiogptunderstandinggeneratingspeech, zhan2024anygpt} primarily focus on enabling multimodal interactions, but these interactions are typically constrained by short context windows and limited reasoning capabilities. Text-to-Audio (TTA) models \cite{kreuk2022audiogen, liu2023audioldm, liu2024audioldm, huang2023make} can generate various audio types, like speech, sound effects, and music. However, since these models prioritize general audio synthesis, they are inherently limited in producing coherent and intelligent spoken content. Although zero-shot Text-to-Speech (TTS) models \cite{casanova2022yourtts,wang2023neural,tan2024naturalspeech} can generate high-quality speech for any speaker, they rely on the provided text and lack the ability to generate long-form informative content.

%A straightforward idea is to combine LLMs with other audio generation models (TTS, TTA) to create a comprehensive framework for podcast production
A straightforward approach is to combine the strengths of these models—using LLMs to generate rich text, TTS models to convert it into spoken content, and TTA models to add appropriate sound effects and background music to produce complete, informative, and professionally structured podcast episodes (Figure \ref{fig:concept} Below). This approach naturally aligns with the emerging paradigm of AI agents \cite{wu2023autogen, langchain}. Empowered by LLMs, various AI agents \cite{ xie2024osworld, du2023improving, lu2024ai} are created to coordinate multiple AI tools to accomplish complex tasks through perception, decision-making, and action execution. A notable implementation of this approach is WavJourney \cite{liu2023wavjourney}, which leverages LLMs to generate an audio script that connects various models for audio generation. While WavJourney represents a significant step forward with its extensive audio generation workflow, its current implementation still faces challenges in producing complete and intellectually rich content (An example demonstrated in Table \ref{tab:wavjourney}).
%WavJourney \cite{liu2023wavjourney} is a well-developed framework designed to create audio programs based on given text descriptions. It leverages LLMs to generate an audio script that connects various models for audio generation. While this work presents a relative extensive audio generation workflow, it still faces limitations in constructing complete and informative content (An example of WavJourney is presented in Table \ref{tab:wavjourney}).

Through observation and analysis of existing automated audio program creation systems, we identify four critical challenges as follows: \textbf{Content Depth and Insight Generation}. For a given topic, how to automatically generate rich and insightful viewpoints and provide meaningful analysis? \textbf{Natural Dialogue Generation}. How to create engaging conversational content that flows naturally between speakers, maintaining coherence while avoiding repetition? \textbf{Appropriate Voice Representation}. How to match suitable voice characteristics with different content and roles, ensuring consistency and authenticity in the audio presentation? \textbf{Speech Quality and Expressiveness}. How to generate robust long-form speech with appropriate prosody and emotion that matches the content's intent and maintains listener engagement?

In this work, we present \model, a fully automated and comprehensive framework for creating content-rich and professionally structured audio programs. To tackle the aforementioned challenges, we:
\begin{itemize}
\item Design a Host-Guest-Writer system that generates engaging and coherent conversation scripts with diverse, insightful viewpoints from various backgrounds and perspectives for any given topic.
%\item Leverage the chain-of-thought (Cot) technique to construct engaging and coherent dialogue content and produce a complete audio generation script step-by-step.
\item Build a preset voice pool through comprehensive voice characteristic analysis to enable dynamic role-voice matching that aligns with speaker personalities and content context.

\item Involve LLM-predicted speaking style in instruction-following TTS model to obtain high-quality speech output with appropriate prosody and emotion.
%(e.g., pitch, timbre, speaking style)
%\item Combining LLM-based style prediction with existing instrucion-following TTS framework to obtain high-quality speech output with appropriate prosody and emotion.
\item Establish comprehensive evaluation metrics for podcast-like audio generation tasks, including assessments of open-ended topic discussions, voice matching, and voice quality.
\end{itemize}

%\model can be used to generate audio programs on any discussion topic. The created content is complete, informative, and well-presented in audio format. Since there is no standard metrics on evaluating such open-ended content generation task, we also provide a benchmark as a reference to evaluate this type of task as thoroughly as possible. Audio samples can be found on the demo page, and the project will be open-sourced to encourage further contributions to this field.
%The generated podcasts are of high quality and ready for publication.

%These instructions are for authors submitting papers to *ACL conferences using \LaTeX. They are not self-contained. All authors must follow the general instructions for *ACL proceedings,\footnote{\url{http://acl-org.github.io/ACLPUB/formatting.html}} and this document contains additional instructions for the \LaTeX{} style files.

%The templates include the \LaTeX{} source of this document (\texttt{acl\_latex.tex}),
%the \LaTeX{} style file used to format it (\texttt{acl.sty}),
%an ACL bibliography style (\texttt{acl\_natbib.bst}),
%an example bibliography (\texttt{custom.bib}),
%and the bibliography for the ACL Anthology (\texttt{anthology.bib}).

\begin{figure*}[t]
  \centering
  \includegraphics[width=\textwidth]{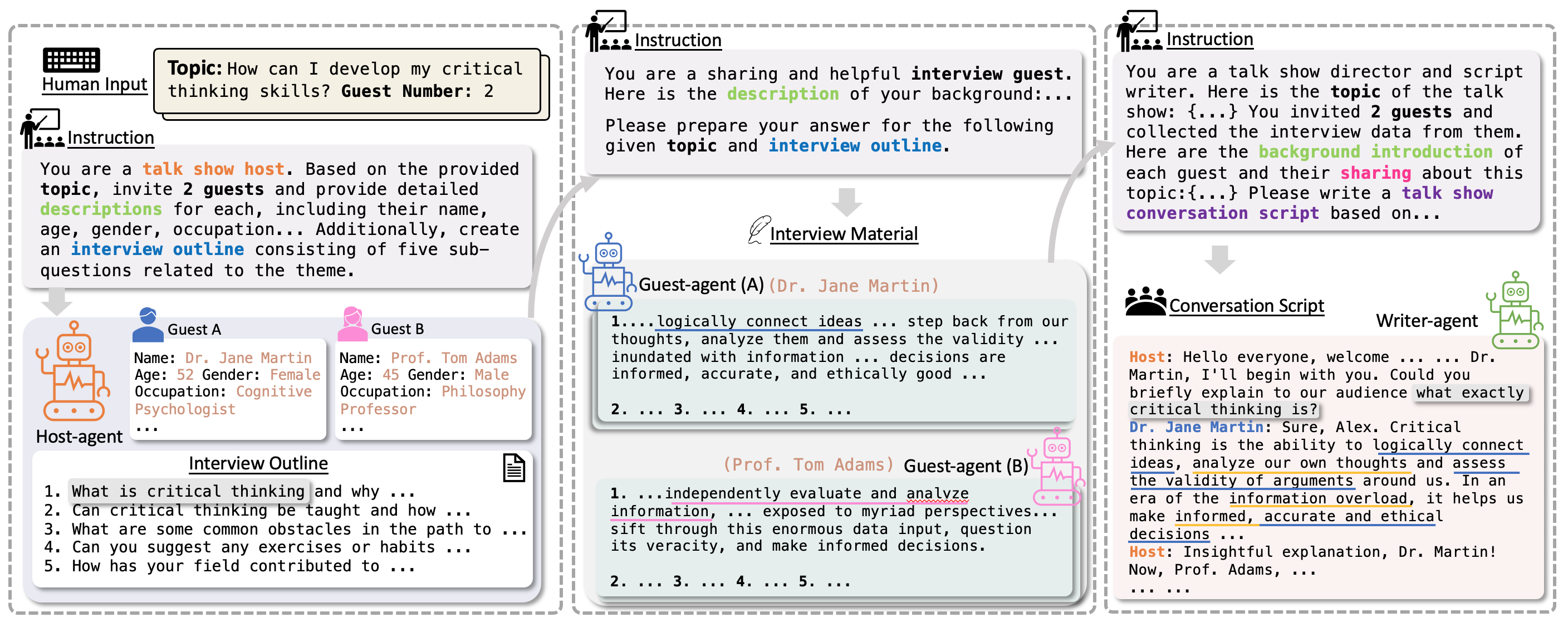}
  \caption{The Workflow of Host-Guest-Writer System. \textbf{Left:} The Host-agent generates guest information and an interview outline based on the given topic and number of guests. \textbf{Middle:} Guest-agents respond to the interview outlines, offering specialized perspectives aligned with their assigned roles. \textbf{Right:} The Writer-agent compiles a complete and coherent conversation script using the gathered interview material.}
  \label{fig:hostguest}
\end{figure*}

\section{Related Works}

\subsection{LLM-powered Agents}
LLMs demonstrate remarkable capabilities in emulating human problem-solving through role-specific configurations \cite{wei2022chain, yao2024tree, shinn2023reflexion}. Building upon this foundation, multi-agent systems incorporate LLMs with diverse role specifications to collaboratively address more complex challenges \cite{liang2023encouraging, talebirad2023multi, chan2023chateval, park2024generative}. Within this framework, each agent functions as a domain expert, focusing on specialized areas and contributing unique perspectives. Furthermore, the integration of multi-modal foundation models has greatly enhanced agents' proficiency in handling cross-modal tasks \cite{huang2023audiogptunderstandinggeneratingspeech, zhang2023speechgpt, hurst2024gpt}. It is crucial to explore effective problem decomposition strategies and appropriate tool utilization for solving real-world problems. 

\subsection{Voice Characteristic Analysis}
%In this work, a voice characteristic analysis tool is required to generate description for the reference voice, which can help with assign an appropriate voice to the speaker in the audio program effectively. 
Voice characteristic analysis is essential in our task for effectively assigning suitable voices to speakers in the audio program. This analysis also known as speech captioning, traditionally relies on approaches that classify and recognize predefined categories from speech signals \cite{issa2020speech}. To address the limitations of insufficient predefined classes, recent studies \cite{yamauchi2024stylecap, xu2023secapspeechemotioncaptioning, zhu2024unistyle} utilize self-supervised learning models for speech feature extraction and description generation. In our analysis, we employ the SpeechCraft \cite{jin2024speechcraft}, an open-source speech dataset with fine-grained text descriptions.

\subsection{Text-to-Speech synthesis}
%Voice content serves as the fundamental component of audio programs for conveying informative messages. Some landmark TTS models \cite{wang2017tacotron, van2016wavenet, ren2020fastspeech, kim2021conditional} have demonstrated the capability to generate human-like voice quality, though they require substantial high-quality speech data for training. 
Recent advances in zero-shot speech synthesis \cite{casanova2022yourtts, wang2023neural, tan2024naturalspeech} enable voice mimicry from a short utterance of a reference speaker. To enhance style control, instruction-following TTS models \cite{yang2024instructtts, guo2023prompttts} bridge textual descriptions with speaking styles. With the growth of the open-source community, an increasing number of outstanding TTS foundation model projects have been released. For instance, Bark \cite{bark2023} is an TTS+ model that extends conventional speech synthesis to include nonverbal cues like laughter, sighs, and crying. CosyVoice \cite{du2024cosyvoice1, du2024cosyvoice} is one of the most recent open-source TTS foundation models, supporting various speech generation tasks like zero-shot voice cloning, multilingual speaking and instruction following.
%offering robust and high-quality speech generation capabilities, 
%However, due to the sensitivity of speech data and the substantial costs associated with data collection, few TTS foundation models are available as open-source for practical applications

%To produce a PDF file, pdf\LaTeX{} is strongly recommended (over original \LaTeX{} plus dvips+ps2pdf or dvipdf). Xe\LaTeX{} also produces PDF files, and is especially suitable for text in non-Latin scripts.

\section{\model}

This section introduces \model, a fully automated comprehensive framework for creating informative and professionally structured audio programs. The focus will be on the key contributions of this work: 1) Host-Guest-Writer system for generating conversation scripts 2) Voice-role matching for selecting suitable voices, and 3) speech synthesis enhanced by LLM-predicted instruction.

\subsection{Host-Guest-Writer system}
%The spoken content serves as the fundamental component of audio programs for conveying informative messages. To generate comprehensive and engaging conversational scripts, we propose a novel Host-Guest-Scriptwriter multi-agent system. The workflow is presented in Figure \ref{fig:hostguest}.
%Following this pattern, the first task of our Host agent is to formulate appropriate guest profiles.
We propose a novel Host-Guest-Writer multi-agent system to generate comprehensive and engaging conversational scripts for audio programs, with the workflow presented in Figure \ref{fig:hostguest}. In real-world talk-show format programs, hosts typically invite several experts to share insights based on their specialized knowledge of the topic. Inspired by this, the first task of our Host-agent is to formulate appropriate guest profiles. Once established, these profiles are assigned to different Guest-agents, enabling them to provide expertise-based responses.
%enabling them to respond based on their designated expertise. 

Rather than implementing computationally intensive turn-by-turn dialogues between agents, which would require managing complex conversation history and turn-taking mechanisms, we adopt a more efficient parallel approach: the Host-agent creates a structured interview outline that serves as a common framework for all Guest-agents to address simultaneously. This allows each Guest-agent to respond to identical questions while maintaining their unique perspectives. Subsequently, all guest responses are processed by a dedicated Writer-agent, which synthesizes the inputs into a cohesive and natural conversational script, effectively eliminating redundancy while preserving the distinct viewpoints of each participant.

In summary, this collaborative framework orchestrates interactions among three specialized agents: the Host-agent, which guides the conversation flow and maintains topic coherence; the Guest-agent, which provides domain expertise and diverse perspectives; and the Writer-agent, which structures and refines the dialogue to ensure natural progression and professional presentation standards.

\subsection{Voice-Role matching}
\begin{figure}[t]
  \includegraphics[width=\columnwidth]{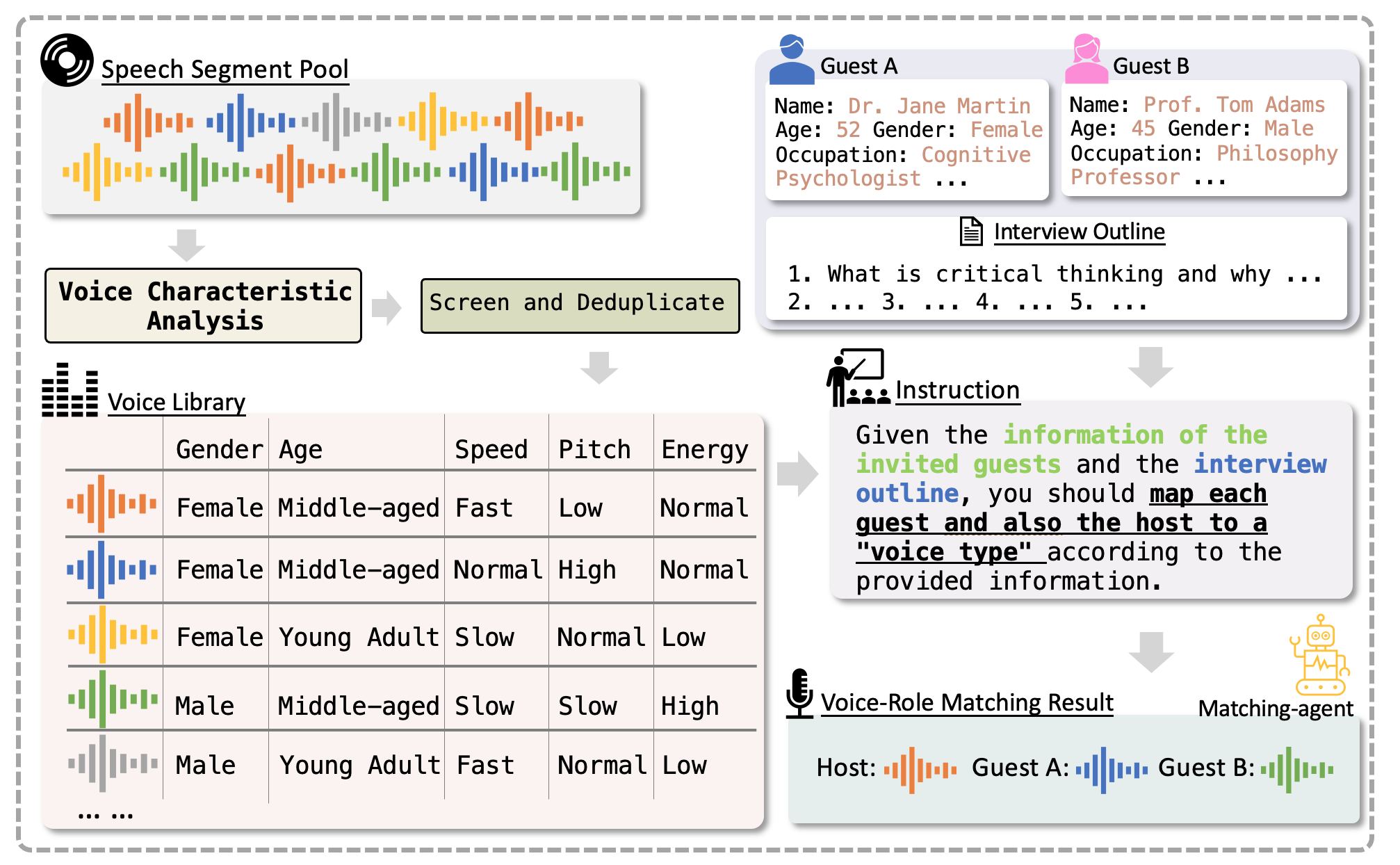}
  \caption{Voice-Role Matching. \textbf{Left:} Construction of a voice library through characteristic analysis of diverse speech segments. \textbf{Right:} The Matching-agent performs voice-role pairing using the voice library, guest profiles, and interview structure.}

  \label{fig:voicematch}
\end{figure}

After obtaining the conversation script, the next crucial step is to select the appropriate voice for each speaker. This voice-role matching process is critical for creating a natural and immersive listening experience for the audience. Figure \ref{fig:voicematch} demonstrates the voice-role matching process. The first step is to collect speech samples from different speakers as much as possible. Then, the collected speech samples will be analyzed to extract voice characteristics for profiling. After that, we screen the profiled speech samples and de-duplicate segments with similar features, ultimately creating a comprehensive and non-redundant voice library. Details of the data can be found in Appendix \ref{sec:appendixv} .
%we apply a voice characteristic analysis tool \footnote{SpeechCraft: https://github.com/thuhcsi/SpeechCraft} to extract characteristics for each speech segment. After that, we screen the speech samples based on their voice characteristics and deduplicate segments with similar features, ultimately creating a comprehensive and non-redundant voice library.

The curated voice library will be provided to the Matching agent, along with the guest information and interview outline generated by the Host-agent. The Matching-agent then leverages all the information to make informed and effective voice-role pairings. This process ensures that the selected voices align naturally with each speaker's designated role and expertise, enhancing the authenticity and engagement of the final audio program.
%The Matching agent then leverage this comprehensive repository of diverse vocal options, coupled with the contextual details about the guest and the interview content, to make informed and effective voice-role pairings. 

\subsection{Instruction-following speech synthesis}
\begin{figure}[t]
  \includegraphics[width=\columnwidth]{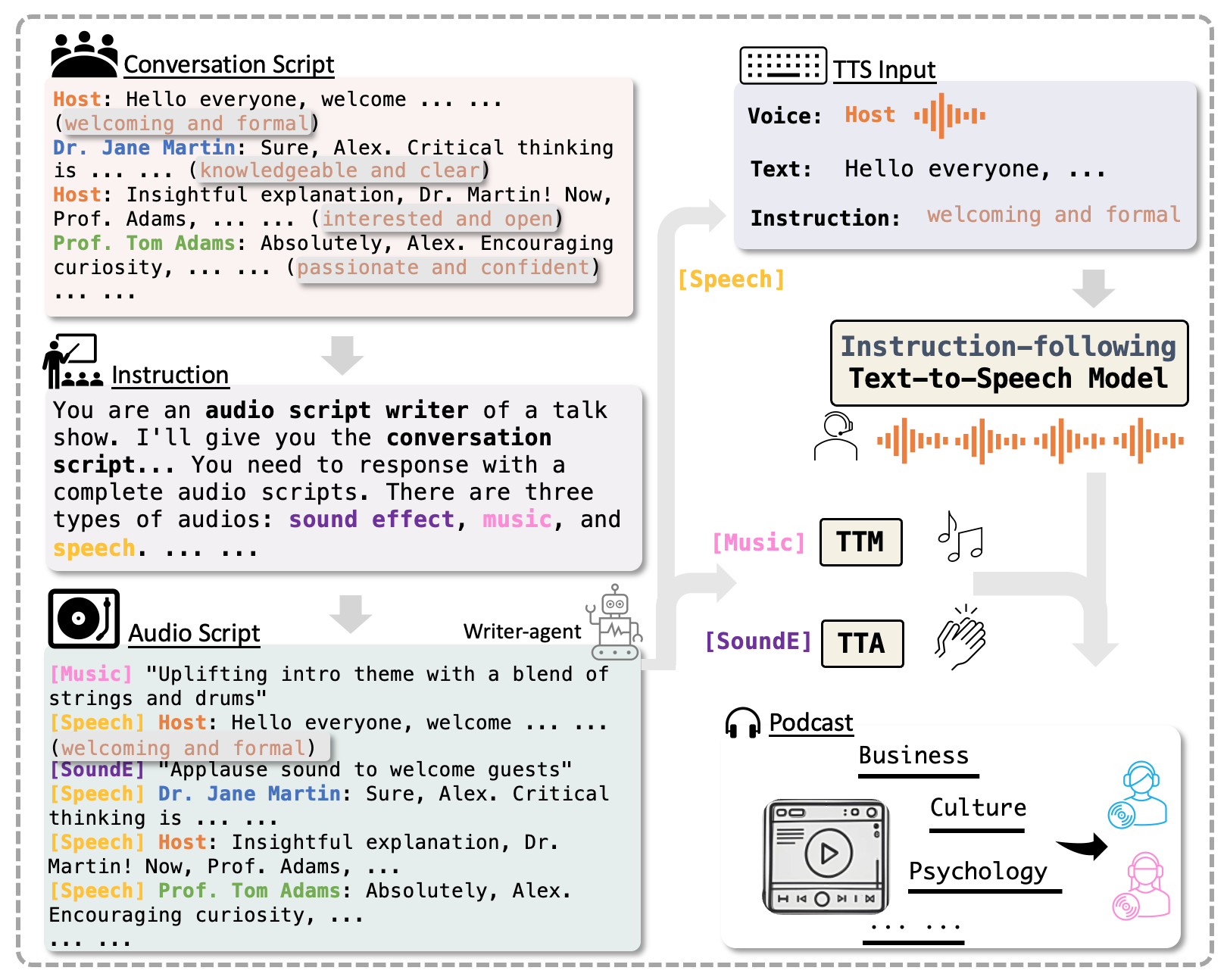}
  \caption{From Conversation Script to Podcast. \textbf{Audio Script Generation:} The Writer-agent create the audio script by enriching the conversation script with sound effect and music. \textbf{Instruction-following TTS:} Speaking styles are generated along with the conversation script, which can be used as instruction to guide the expressive speech synthesis. \textbf{Audio Production:} The generated audio segments are combined to create the final podcast.}

  \label{fig:audio}
\end{figure}

To enhance the expressiveness of generated speech, we leverage LLM-predicted speaking styles as instructions to guide the synthesis. As depicted in the top right of Figure \ref{fig:audio}, the instruction-following TTS system takes three inputs: the text (content to be spoken), a reference voice (speech segment), and an instruction (speaking style). The system then generates speech in the voice of the reference speaker, adhering to the specified speaking style.

%The first line of the file must be
%\begin{quote}
%\begin{verbatim}
%\documentclass[11pt]{article}
%\end{verbatim}
%\end{quote}

%To load the style file in the review version:
%\begin{quote}
%\begin{verbatim}
%\usepackage[review]{acl}
%\end{verbatim}
%\end{quote}
%For the final version, omit the \verb|review| %\begin{quote}
%\begin{verbatim}
%\usepackage{acl}
%\end{verbatim}
%\end{quote}

%To use Times Roman, put the following in the preamble:
%\begin{quote}
%\begin{verbatim}
%\usepackage{times}
%\end{verbatim}
%\end{quote}
%(Alternatives like txfonts or newtx are also acceptable.)

%Please see the \LaTeX{} source of this document for comments on other packages that may be useful.

%Set the title and author using \verb|\title| and \verb|\author|. Within the author list, format multiple authors using \verb|\and| and \verb|\And| and \verb|\AND|; please see the \LaTeX{} source for examples.

%By default, the box containing the title and author names is set to the minimum of 5 cm. If you need more space, include the following in the preamble:
%\begin{quote}
%\begin{verbatim}
%\setlength\titlebox{<dim>}
%\end{verbatim}
%\end{quote}
%where \verb|<dim>| is replaced with a length. Do not set this length smaller than 5 cm.

\section{Experimental Setups}
Our experiments will center around the key contributions of this work: topic-based discussion content generation, voice-role matching, and expressive speech synthesis. We follow WavJourney's setup for generating music and sound effects but use a more recent open-source framework, CosyVoice2,  for speech generation. 
%As a more recent open-source framework, CosyVoice2 is more robust, especially for long-form generation.

\subsection{Datasets}
We base our evaluation on a subset of data from \cite{chiang2023vicuna}, which originally contains 80 questions across 8 categories. To align with the topic-discussion scenario, we exclude categories such as "code" and "match" that are less suitable, and ultimately select 4 categories, Generic, Knowledge, Common-sense, and Counterfactual. This resulted in 40 topics, with 10 topics per category, serving as our experimental data. The experimental data is English-based. In addition to it, we also showcase some Chinese-based podcasts on the demo page to provide a broader perspective on \model's capabilities.

%Topic-based discussion content
\subsection{Evaluation on conversation scripts}
The dialogue content in podcast programs typically revolves around a given theme, which can vary widely to cater to different audience interests. The discussions typically showcase participants' unique perspectives and personal insights, offering listeners a rich tapestry of viewpoints and thought-provoking ideas. Given this subjective and open-ended nature of podcast content, establishing definitive ground truth or applying standardized quality metrics becomes particularly challenging.

To address this, we design the evaluation methods from two aspects. First, We employ several \textbf{quantitative metrics} that measure the lexical diversity, semantic richness and information density of the generated content. These metrics operate independently of any reference or ground truth text, focusing solely on the characteristics of the text itself. Second, we utilize \textbf{LLM-as-a-Judge} methodology to perform comparative quality assessments between discussion texts. The specific implementation details of both approaches are outlined below.
% to provide a certain degree of objective quality assessment of the dialogue content. 

\begin{figure}[t]
  \includegraphics[width=\columnwidth]{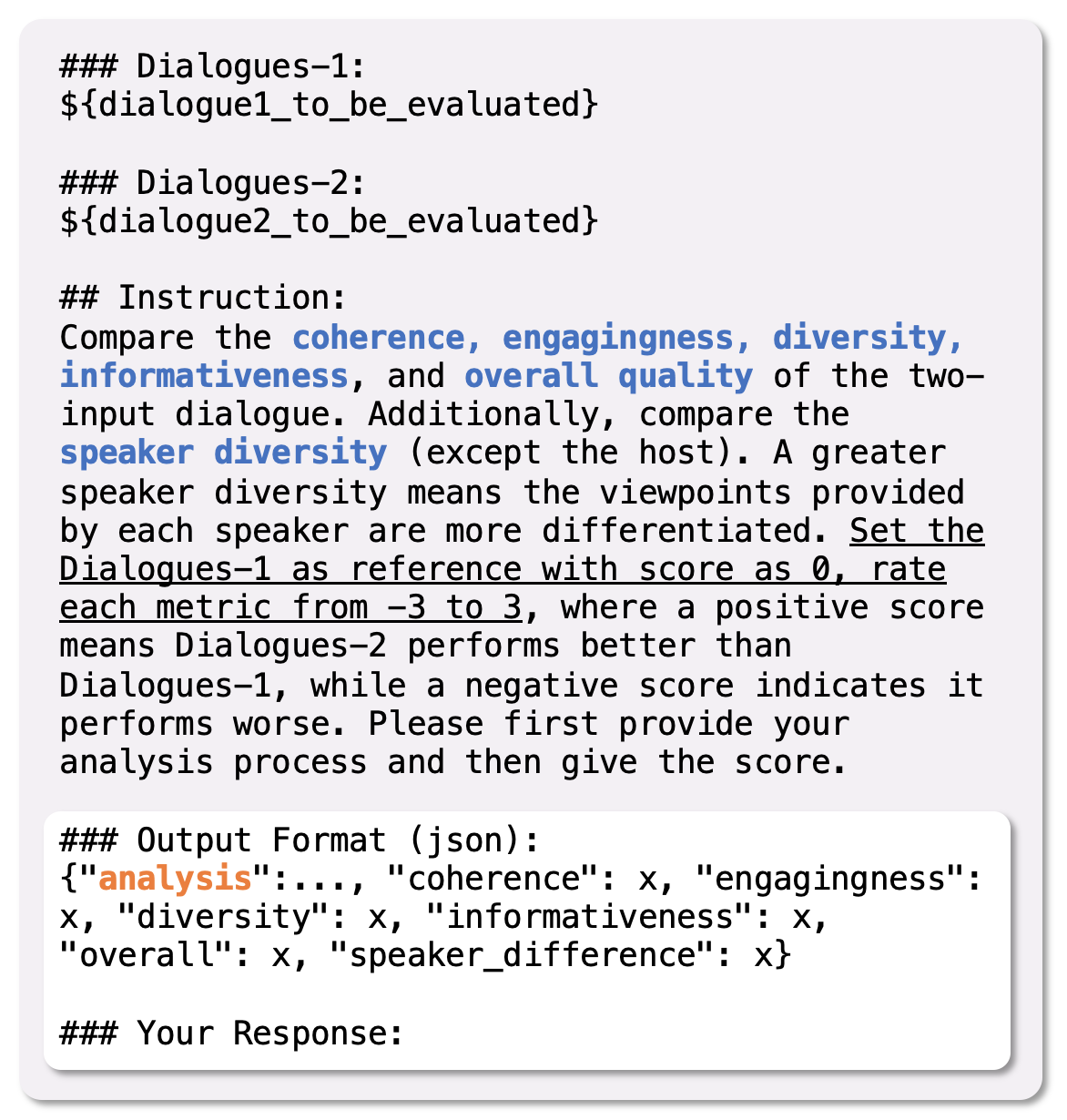}
  \caption{Prompt for GPT-4 Evaluator.}

  \label{fig:gpteval}
\end{figure}

\begin{itemize}[left=0pt]

\item \textbf{Distinct-N} \cite{li2015diversity}, particularly distinct-1 and 2, is used to evaluate the text diversity. It emphasizes the count of distinct n-grams within the text, thereby penalizing those that contain many repeated words. To ensure comparability between texts of varying lengths, we employ a sliding window for normalization of scores. The window size is set as 100 and similar normalization are applied to other quantitative metrics.
\begin{equation}
\text{Distinct-N} = \frac{1}{N_w} \sum_{i=1}^{N_w} \frac{| \text{UniqueNgrams}_i |}{| \text{TotalNgrams}_i |}
\end{equation}
where $N_w$ is the number of sliding windows.

\item \textbf{Semantic-Div} This metric is measured by calculating the cosine distance between text segments using BERT \cite{devlin2018bert} embeddings, providing a robust measure of semantic diversity. 
\begin{equation}
\text{Semantic-Div} = \text{mean}\left(1 - \cos\left(\mathbf{e}_i, \mathbf{e}_j\right)\right)
\end{equation}

$\mathbf{e}_i$ and $\mathbf{e}_j$ are BERT embeddings of different text windows.
The cosine similarity between the embeddings is calculated as $\cos\left(\mathbf{e}_i, \mathbf{e}_j\right) = \frac{\mathbf{e}_i \cdot \mathbf{e}_j}{|\mathbf{e}_i| |\mathbf{e}_j|}$.
%and then subtracted from 1 to obtain the distance. The mean of all the pairwise distances between the text window embeddings is taken as the final Semantic Diversity score.

%Vocabulary Richness
\item \textbf{MATTR} \cite{covington2010cutting} is for measuring lexical diversity by calculating the average Type-Token Ratio (TTR) across a sliding window. This approach reduces sensitivity to text length, providing a robust measure of vocabulary richness. Unlike metrics like Distinct-N, which focus on n-gram diversity, MATTR emphasizes word-level diversity, which useful for analyzing the linguistic richness of natural texts.
%Type-Token Ratio (TTR), which is the ratio of unique tokens to total tokens, is a simple measure of lexical diversity. Since TTR is affected by the length of the text sample, we use Moving-Average TTR (MATTR) \cite{covington2010cutting}, which is the average of the TTR values computed within a sliding window of the text. 

\begin{equation}
\text{MATTR} = \frac{1}{N_w} \sum_{i=1}^{N_w} \text{TTR}_i
\end{equation}

%Information Density
%over token distributions, excluding stopwords
\item \textbf{Info-Dens} We measure information density using Shannon entropy \cite{shannon1948mathematical}:
\begin{equation}
\text{Info-Dens} = -\sum_{i=1}^{N} p_i \log_2 p_i
\end{equation}

where $N$ is the number of unique tokens in the filtered token sequence (excluding stopwords),  $p_i$ is the probability of token i in the filtered sequence.

\item \textbf{LLM-as-a-Judge} The emergence of LLMs as evaluation tools \cite{zheng2023judging} has revolutionized assessment methods for open-ended content generation. This approach proves particularly valuable when dealing with tasks lacking definitive answers and where traditional manual evaluation would be resource-intensive and potentially subjective. We design the evaluation prompt as shown in Figure \ref{fig:gpteval}. The key \underline{design principles} are: \textcolor{blue}{1)} The evaluation metrics are primarily based on the template presented in \cite{zhang2024comprehensiveanalysiseffectivenesslarge} for prompting GPT-4 to annotate the dialogue data. Additionally, we introduce a new metric, speaker diversity, to assess the diversity of viewpoints between different speakers. \textcolor{blue}{2)} We opt for a comparative evaluation between two samples, allowing the scores to reflect the relative quality of the dialogues. \textcolor{blue}{3)} To address the potential issue of position bias mentioned in \cite{zheng2023judging}, we conduct two evaluations for each pair of dialogues - one with "dialogue A vs dialogue B", and another with "dialogue B vs dialogue A". We then average the results to obtain a more fair score. \textcolor{blue}{4)} Drawing from the approach described in \cite{wang2023large}, we first ask the LLM to generate an explanation (evaluation evidence), and then provide the score. This allows the score to be calibrated with the evaluation evidence. \textcolor{blue}{5)} We use GPT-4 as the evaluator, ensuring more robust and reliable results.

\end{itemize}

%Voice-Role matching
\subsection{Evaluation on voices}
\textbf{Voice-Role matching} To evaluate the effectiveness of \model's voice-role matching mechanism, we conduct a subjective perception study involving 6 participants from diverse backgrounds. The study requires participants to evaluate 40 generated podcast segments, assessing whether the voices of the host and guests are appropriate and coherent with their assigned roles and the discussion topics. The participants provide a binary "pass" or "fail" judgment for each speaker in each segment. The overall pass rate serves as the key metric to quantify the success of the voice-role matching process.

\noindent \textbf{Instruction-following speech synthesis} To assess the impact of LLM-predicted speaking styles in \model, we employ two evaluation metrics: preference scores and comparative mean opinion scores (CMOS). Both evaluation metrics are obtained by comparing speech generated with and without instruction guidance. For preference score, 9 evaluators selected between three options for each pair: ``A wins'', ``No preference'', or ``B wins''. The CMOS require judgers to rate the sample pairs on a scale from -3 to 3, with the non-instructed sample as a reference point at 0. The audio samples are speech segments from the generated podcasts.

\section{Experimental Analysis}
Throughout our experiments, we set the guest number to 2, except for those conducted in the guest number analysis. To ensure experimental consistency and reliable performance, we exclusively employed GPT-4 for all LLM-dependent tasks.

%Host-Guest-Writer System
\subsection{Analysis on conversation scripts}
To evaluate the effectiveness of our Host-Guest-Writer system, we conduct a comparative assessment against a baseline approach. The baseline implementation utilizes GPT-4 to directly generate conversation scripts from given topics, using the following prompt structure:

\textit{You are a talk show director and script writer. Here is the topic of the talk show: ... Please Write a corresponding talk show conversation script featuring 1 host and 2 guests. }

\noindent We chose not to use WavJourney as a baseline due to the dialogue scripts it generates (Table \ref{tab:wavjourney}) are very short and significantly lower in quality compared to our method. While dialogue scripts generated directly by GPT-4 (Table \ref{tab:directly}) are obvious content-rich than those created by WavJourney. 

Table \ref{tab:multiagent} presents our comparative evaluation results between the proposed system and the baseline. The quantitative metrics are expressed as difference scores ranging from -2 to 2, calculated by subtracting baseline scores from our system's scores. Additionally, we employed the "LLM-as-a-Judge" methodology, utilizing GPT-4 to provide comparative assessments on a scale of -3 to 3. For both evaluation approaches, positive values indicate superior performance by our proposed system, while negative values favor the baseline. Detailed metric descriptions can be found in Section 4.2.

The results demonstrate consistent and substantial improvements across all evaluation dimensions for conversations generated by the Host-Guest-Writer system. With only one minor exception - a marginal decline of -0.005 in the Semantic-Div score for the Generic category - our system outperformed the baseline across all metrics. These comprehensive positive results strongly validate the effectiveness of our Host-Guest-Writer approach in generating high-quality conversational content.

\begin{table*}[htbp]
\centering
\setlength{\tabcolsep}{6pt}
\renewcommand{\arraystretch}{1.2}
\begin{tabular}{|c|c|c|c|c|c|}
\hline
\multicolumn{2}{|c|}{\diagbox{\textbf{Metrics}}{\textbf{Categories}}} & \textbf{Generic} & \textbf{Knowledge} & \textbf{Common-sense} & \textbf{Counterfactual} \\ 
\hline
\multirow{5}{*}{\textbf{Quantitative Metrics}} 
& \cellcolor{blue!10}Distinct\_1 & \cellcolor{blue!10}+0.031 & \cellcolor{blue!10}+0.034 & \cellcolor{blue!10}+0.028 & \cellcolor{blue!10}+0.005 \\ 
& \cellcolor{blue!10}Distinct\_2 & \cellcolor{blue!10}+0.016 & \cellcolor{blue!10}+0.008 & \cellcolor{blue!10}+0.011 & \cellcolor{blue!10}+0.004 \\ 
& \cellcolor{blue!10}Info-Dens & \cellcolor{blue!10}+0.707 & \cellcolor{blue!10}+0.705 & \cellcolor{blue!10}+0.670 & \cellcolor{blue!10}+0.558 \\ 
& \cellcolor{blue!10}Semantic-Div & \cellcolor{blue!10}\underline{-0.005} & \cellcolor{blue!10}+0.010 & \cellcolor{blue!10}+0.019 & \cellcolor{blue!10}+0.008 \\ 
& \cellcolor{blue!10}MATTR & \cellcolor{blue!10}+0.031 & \cellcolor{blue!10}+0.034 & \cellcolor{blue!10}+0.028 & \cellcolor{blue!10}+0.005 \\ 
\hline
\multirow{6}{*}{\textbf{LLM-as-a-Judge}}
& \cellcolor{yellow!10}Coherence & \cellcolor{yellow!10}+0.7000 & \cellcolor{yellow!10}+0.6500 & \cellcolor{yellow!10}+0.6500 & \cellcolor{yellow!10}+0.6500 \\ 
& \cellcolor{yellow!10}Engagingness & \cellcolor{yellow!10}+1.4500 & \cellcolor{yellow!10}+1.4500 & \cellcolor{yellow!10}+1.4000 & \cellcolor{yellow!10}+1.2500 \\ 
& \cellcolor{yellow!10}Diversity & \cellcolor{yellow!10}+1.6500 & \cellcolor{yellow!10}+1.2500 & \cellcolor{yellow!10}+1.5500 & \cellcolor{yellow!10}+1.0000 \\ 
& \cellcolor{yellow!10}Informativeness & \cellcolor{yellow!10}+1.9000 & \cellcolor{yellow!10}+1.8000 & \cellcolor{yellow!10}+2.2000 & \cellcolor{yellow!10}+1.7000 \\ 
& \cellcolor{yellow!10}Speaker-diversity & \cellcolor{yellow!10}+1.1500 & \cellcolor{yellow!10}+1.1000 & \cellcolor{yellow!10}+1.3000 & \cellcolor{yellow!10}+1.1000 \\ 
& \cellcolor{yellow!10}Overall & \cellcolor{yellow!10}+1.7500 & \cellcolor{yellow!10}+1.6625 & \cellcolor{yellow!10}+1.7250 & \cellcolor{yellow!10}+1.4000 \\ 
\hline
\end{tabular}
\caption{Evaluation on the Host-Guest-Scriptwriter System. \textbf{Baseline:} Directly ask the GPT-4 to generate a conversation script for a provided topic. \textbf{Quantitative metrics:} derived by subtracting the baseline score from the proposed model's score, yielding a range of -2 to 2. \textbf{LLM-as-a-Judge} scores range from -3 to 3. Positive values in all metrics indicate that the proposed model outperforms the baseline, whereas negative values suggest the opposite. }
\label{tab:multiagent}
\end{table*}

%Voice-role matching
\subsection{Analysis on voices}
\begin{figure}[t]
  \includegraphics[width=\columnwidth]{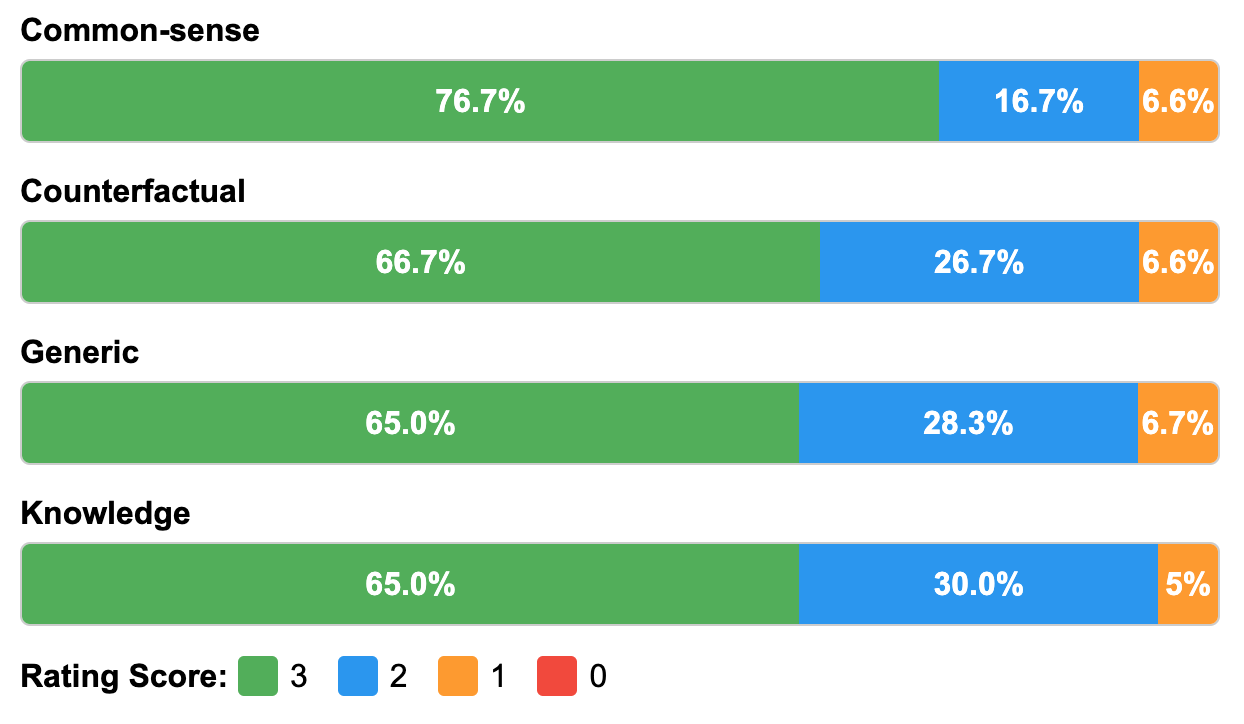}
  \caption{Voice-Role Matching result. This perception test is designed to evaluate whether the tone of each voice actor aligns with their respective roles and scenarios. The rating scale ranges from 0 to 3, where the score indicates the number of speakers that match effectively.}

  \label{fig:match_exp}
\end{figure}

\textbf{Voice-Role matching} Figure \ref{fig:match_exp} illustrates the Voice-Role matching evaluation results. With setup of two guests plus one host (three speakers total), a rating scale of 0-3 is to indicate the number of speakers successfully matching their assigned roles. The findings demonstrate robust performance across all categories, with over 60\% of sessions achieved full matches where all voices aligned with their roles. Moreover, more than 90\% of sessions achieved successful matching for at least two speakers. The pass rates for the categories Common-sense, Counterfactual, Generic, and Knowledge are 90.0\%, 86.7\%, 86.1\%, and 86.7\%, respectively.

\begin{figure}[t]
  \includegraphics[width=\columnwidth]{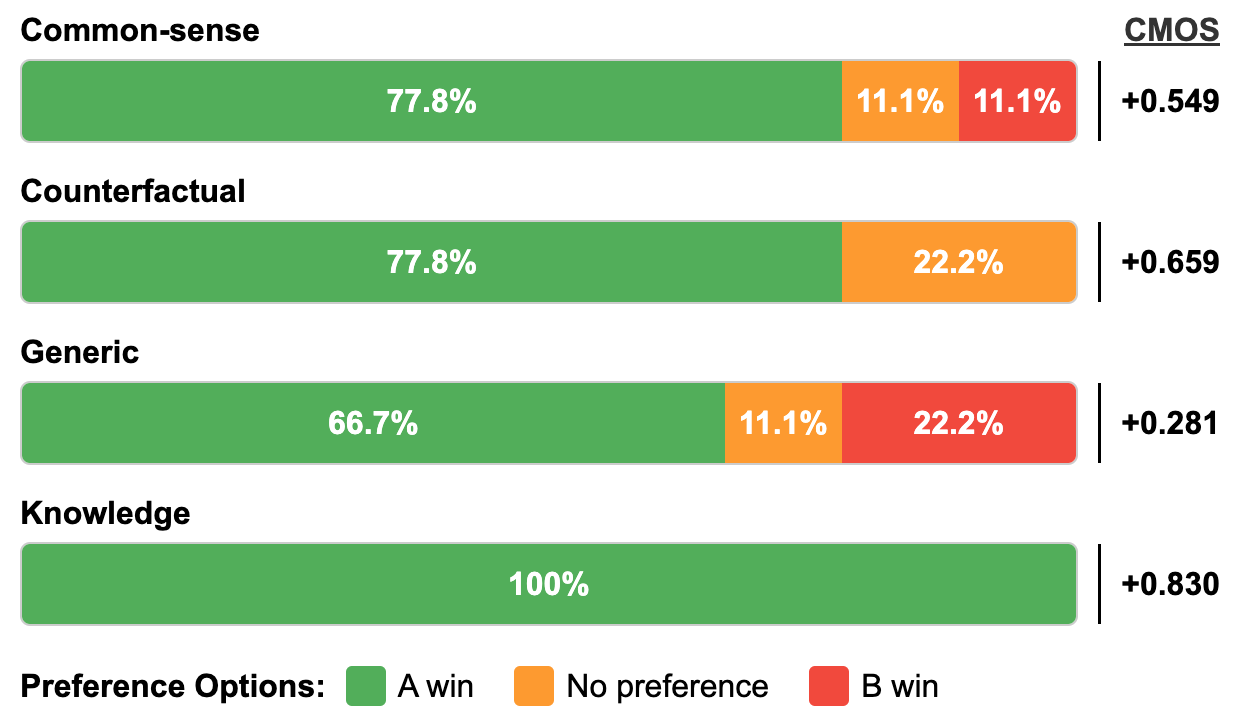}
  \caption{Evaluation on the instruction-following TTS. \textbf{Preference Test:} A - speech generation with guidance of LLM-predicted speaking style; B - speech generation without it.  \textbf{CMOS Test:} Set B as reference, rating A from -3 to 3. Positive means A is better.}

  \label{fig:cmos}
\end{figure}

\noindent \textbf{Instruction-following speech synthesis} Figure \ref{fig:cmos} showcases the preference and the CMOS scores. The analysis demonstrates a clear preference for speech samples generated using LLM-predicted speaking styles across all categories. Furthermore, all CMOS scores are positive, ranging from 0.2 to 0.9, further supporting this conclusion.

\begin{table*}[htbp]
\centering
\begin{tabular}{@{}>{\hspace{1.5em}}l|ccccc@{}}
\toprule
\diagbox[width=10em,height=3em,innerleftsep=1.5em]{Methods}{Metrics} & \textbf{Distinct\_1} & \textbf{Distinct\_2} & \textbf{Info-Dens} & \textbf{Semantic-Div} & \textbf{MATTR} \\ \midrule
\#Guest = 1       & 0.7392               & 0.9768               & 7.1247             & 0.1521                 & 0.7392         \\
\#Guest = 2       & \textbf{0.7662}      & \textbf{0.9789}      & 7.0971             & \textbf{0.2310}        & \textbf{0.7662} \\
\#Guest = 3       & 0.7367               & 0.9767               & \textbf{7.1323}    & 0.1230                 & 0.7367         \\
\#Guest = 4       & 0.7278               & 0.9768               & 6.9253             & 0.0977                 & 0.7278         \\
\#Guest = 5       & 0.7012               & 0.9641               & 7.1328             & 0.1150                 & 0.7037         \\ \midrule
\#Guest = 2 (w/o outline) & 0.7037       & 0.9569               & 6.9515             & 0.1275                 & 0.7037         \\ \midrule
\#Guest = 2 (Single Agent) & 0.6559      & 0.9303               & 5.9001             & 0.1277                 & 0.6559         \\ \bottomrule
\end{tabular}
\caption{Ablation Study. 1. Guest Number; 2. with or without preset outline; 3. multi-agent VS single-agent }
\label{tab:ablation}
\end{table*}

\subsection{Ablation Study}
We perform ablation studies on our Host-Guest-Writer system to verify the effectiveness of three critical factors: guest number, outline, and multi-agent framework. Table \ref{tab:ablation} presents the results. 
%We conduct an ablation study on the host-guest-writer system

\noindent \textbf{Guest number} We explore the influence of the guest number by varying it from 1 to 5. Our findings indicate that increasing the number of participants does not always enhance content quality. We found that 2-guest setups consistently produce the most informative scripts based on quantitative metrics. While larger groups offer diverse perspectives, they often introduce redundancy and coordination challenges. Smaller groups foster more focused interactions and deeper dialogue, allowing participants to fully develop their ideas. 
%This aligns with the principle that meaningful conversation quality often trumps participant quantity.
%We explore the influence of the guest number by varying it from 1 to 5. Our findings indicate that increasing the number of participants does not always enhance content quality. Quantitative metrics show that the conversation scripts generated by 2 guests deliver the most informative content. While a larger group can bring in more ideas, it often leads to confusion and redundancy. A few insightful individuals can create deeper dialogue than a big crowd. Smaller groups typically encourage better interaction and effective expression, whereas managing time becomes more challenging as the number of participants increases. 

\noindent \textbf{Outline} The use of a topic-centered outline, created by the "Host", serves as a crucial structural framework for guiding guest interactions. Our experiments compare performance between scenarios with and without this outline, using two guests in both cases. Results show that \textit{\#Guest = 2 (w/o outline)} performs worse than \textit{\#Guest = 2} with outline, demonstrating the importance of structured guidance in the Host-Guest-Writer system.
%One of the tasks of the "Host" is to create a topic-centered outline. This preset outline will be used for each guest, helping to organize the entire conversation. By removing the preset outline, \textit{\#Guest = 2 (w/o outline)} shows a decline in performance compared to \textit{\#Guest = 2} with the outline. This illustrates the importance of structured guidance in the host-guest-writer system.

\noindent \textbf{Multi-agent system} Our Host-Guest-Writer system is a multi-agent framework that collaborate multiple LLMs with distinct role settings to compose insightful conversation scripts from diverse perspectives. To evaluate its effectiveness, we compare it against a single-agent approach where one LLM handles all tasks. The comparison baseline \textit{\#Guest = 2 (Single Agent)} receives the instruction:
%Our \textbf{multi-agent} host-guest-writer system leverages multiple LLMs with distinct roles to generate rich, multi-perspective conversation scripts. To evaluate its effectiveness, we compare it against a single-agent approach where one LLM handles all tasks. The comparison baseline \textit{#Guest = 2 (Single Agent)} receives the following instruction:

\textit{You are a talk show director and script writer. Here is the topic of the talk show:... Please follow the steps: 1. Based on the provided topic, invite 2 guests and provide detailed descriptions for each, including their ... 2. Create an interview outline consisting of five sub-questions related to the theme. 3. Write a talk show conversation script based on the unique role, experiences and diverse perspective of each invited guest...}

\noindent As evidenced in Table \ref{tab:ablation}, the multi-agent collaborative system demonstrates clear performance advantages over the single-agent approach across all evaluated metrics.
%From the result showed in Table \ref{tab:ablation}, it is obvious that a multi-agent collaborating system outperforms the single-agent system.

\subsection{Case Study}
In the Appendix \ref{sec:appendixc}, we provide comparative examples of conversation scripts on the topic: ``How can I develop my critical thinking skills?'' Table \ref{tab:wavjourney} features dialogue content extracted from the audio script generated by WavJourney, which is notably short and lacks depth, with only 4 topic-related exchanges and providing limited information. This is due to that WavJourney generates dialogue as part of an audio script, which restricts multi-turn discussions. Table \ref{tab:directly} presents scripts generated by directly asking GPT-4 with baseline instruction presented in Section 5.1. It shows modest improvement but still constrained to 4 turns. Table \ref{tab:singleagent} showcases the Single-Agent version of the Host-Guest-Writer system we discussed in section 5.3. This case achieves richer content through task decomposition but lacks concluding remarks. Table \ref{tab:casestudty_m} displays the conversation scripts produced by our proposed \model's Multi-Agent Host-Guest-Writer system, delivering the most comprehensive and well-structured discussion of the topic.
% Table 5 demonstrates the Single-Agent Host-Guest-Writer system, which achieves richer content through task decomposition but lacks concluding remarks. Table 6 features our proposed \model's Multi-Agent Host-Guest-Writer system, delivering the most comprehensive and well-structured discussion of the topic.

\section{Conclusion}
In this study, we proposed \model, a comprehensive framework for creating audio programs that addresses the shortcomings of previous automated podcast-like generation methods. The key components of \model\ include: 1) a Host-Guest-Writer system generating comprehensive, multi-perspective conversation scripts, 2) a preset diverse voice pool for suitable voice-role assignment, and 3) LLM-guided speech generation for enhanced expressiveness. Given the absence of established benchmarks in podcast generation, we designed a thorough experimental setup encompassing both qualitative and LLM-based evaluation of the conversation scripts, as well as voice-related metrics. Our extensive experimental results demonstrate \model's capability to produce high-quality, complete, and realistic audio programs.
% This paper introduces \model, a novel framework that overcomes limitations in existing automated podcast generation systems. Our approach comprises three key innovations: 1) a Host-Guest-Writer system generating comprehensive, multi-perspective conversation scripts, 2) a curated voice pool enabling context-appropriate voice-role matching, and 3) LLM-guided speech generation for enhanced expressiveness. Given the absence of established benchmarks in podcast generation, we developed a comprehensive evaluation framework incorporating qualitative analysis, LLM-based assessment of conversation content, and voice quality metrics. Our extensive experimental results demonstrate \model's capability to produce engaging, complete, and naturalistic audio programs.
\section{Discussion}
\textbf{Limitations} Although \model\ is the first fully-automatic system capable of generating complete and informative podcast-like audio, several limitations in this study require further investigation: 1) Voice Quality. While we used a state-of-the-art open-source TTS foundation model to generate speech, offering improved robustness compared to earlier models, quality issues may still arise when generating large amounts of long-form content. 2) Voice Pool. In this work, reference speech segments were collected from LibriTTS \cite{koizumi2023libritts}. To produce more natural conversational audio, it is essential to expand the voice pool by incorporating more conversational-style voices. 3) Sound Effects and Music. This study primarily focuses on improving content and voice generation. However, there is room for enhancement in generating sound effects and music, as well as determining their appropriate placement within the audio.

%Although \model should be the first work that can generate complete and informative podcast-like audio, there are still several limitations in this study can be further inverstigation: 1) Voice quality. Although we have used the most recent open-source TTS foundation model to generate speech, which has better robustness compared with most previous models, it still may encounter quality issues when generating amount of long-form content. 2) Voice pool. We collect reference speech segments from LibriTTS \cite{koizumi2023libritts} in this work. For more natural conversation generation, it is enssential to collect more conversational-style voices into the voice pool. 3) Sound effect and music. This works focus on the improving of content and voice generation, the generation of sound effect and music and find the approporiate position to insert also has space to be improved.  

\noindent \textbf{Future Work} To improve the podcast-listening experience, beyond just expanding the voice library diversity, a more advanced approach would be to generate new synthetic voices directly based on the desired characteristics, which can be more intelligent and help avoid some of the ethical concerns around real-voice cloning and consent. Additionally, the conversational expression can be further enhanced by adopting a more casual and natural style and incorporating appropriate vocal articulations like laughter, sighs, exclamations, and other non-semantic vocalizations, as incorporating these expressive sounds can make the conversation feel more lively and engaging for the user. 

\section{Ethics Statement}
Since this work involves generating long-form audio content, including speech, music, and sound effects, we address considerations as follows: 1) Copyright and Intellectual Property. The generation of music, sound effects, and voices must respect existing copyright laws. In this work, we rely on open-source datasets and models to ensure compliance with intellectual property rights. Users are encouraged to verify that their use of \model\ complies with copyright regulations. 2) Voice Cloning and Consent. In this study, we use anonymized, open-source speech data to avoid ethical violations. Users must ensure they have proper authorization and consent when using the system to generate speech resembling real individuals.

\bibliography{acl_latex}

\onecolumn
\appendix

\section{Voice Library}
\label{sec:appendixv}
\textbf{English} For the voice pool construction detailed in Section 3.2, we utilize speech segments from the LibriTTS-R dataset \cite{koizumi2023libritts}, an enhanced sound-quality version of the LibriTTS corpus \cite{zen2019libritts}. This dataset contains approximately 350,000 speech segments from over 2,000 speakers, with a balanced gender distribution. By screening and eliminating duplicate segments with similar voice characteristics, we develop a diverse voice library containing 222 unique speakers.

\noindent \textbf{Madarin} We also construct a Mandarin voice pool using the AISHELL-3 speech corpus \cite{shi21c_interspeech}, which contains approximately 85 hours of recordings from 218 native Mandarin speakers, totaling over 8,000 utterances. We do selection and deduplication based on the voice characteristic labels provided by SpeechCraft \cite{jin2024speechcraft}. As a result, we curated a diverse voice library comprising 172 speech segments from 85 distinct speakers.

\section{Case Study}
\label{sec:appendixc}
Please refer to the examples shown in the tables below and the illustration provided in Section 5.4.

\begin{table*}[htbp]
\centering
\begin{tabular}{|>{\raggedright\arraybackslash}p{3cm}|>{\raggedright\arraybackslash}p{12cm}|}
\hline
\textbf{Speaker} & \textbf{Speaking Content} \\ \hline
Host & Hello everyone, and welcome to our show. Today, we're discussing a very intriguing topic: How can I develop my critical thinking skills? \\ 
  & To shed light on this topic, I'm delighted to introduce our first guest. \\ 
  & Please welcome, Dr. Jane Doe, a renowned psychology professor. \\ \hline
Jane Doe & Thank you for having me here. I'm excited to share a few insights on improving critical thinking skills. \\ \hline
Host & Fantastic, we also have another guest joining us for the discussion. Please welcome, Mr. John Smith, a successful entrepreneur. \\ \hline
John Smith & Thank you. I'm thrilled to be here to discuss this important topic. \\ \hline
\cellcolor{blue!10}Host & \cellcolor{blue!10}Let's dive right into it then. \\ 
\cellcolor{blue!10}  & \cellcolor{blue!10}Firstly, Dr. Doe, could you please explain what is critical thinking? \\ \hline 
\cellcolor{yellow!10}Jane Doe & \cellcolor{yellow!10}Sure. Critical thinking refers to the ability to analyze information objectively and make a reasoned judgment. \\ \hline
\cellcolor{blue!10}Host & \cellcolor{blue!10}Excellent! Now, Mr. Smith, how has critical thinking played a role in your success? \\ \hline
\cellcolor{green!10}John Smith & \cellcolor{green!10}Critical thinking has been essential in my decision-making process. It's helped weigh my options logically before making a decision. \\ \hline
Host & Thank you both for your contributions. I believe this discussion helped our audience understand more about critical thinking. \\ \hline
\end{tabular}
\caption{Example of conversation scripts generated by \textbf{WavJourney}. Generate a talk show featuring one host and two guests. The topic is: How can I develop my critical thinking skills?} % Add a caption for the table
\label{tab:wavjourney} % Add a label for referencing the table
\end{table*}

\begin{table*}[htbp]
\centering
\begin{tabular}{|>{\raggedright\arraybackslash}p{3cm}|>{\raggedright\arraybackslash}p{12cm}|}
\hline
\textbf{Speaker} & \textbf{Speaking Content} \\ \hline
Host & Hello and welcome to 'Mind Matters'! I'm your host, Steve, and I'm thrilled to have you join us for another exciting episode. Tonight, we delve into the human mind and explore the topic of critical thinking skills. \\ 
  & Joining us tonight are two distinguished guests, Dr. Linda, a cognitive psychologist, and James, a renowned author of self-improvement literature. Welcome to the show, Dr. Linda and James! \\ \hline
Dr. Linda & Thank you, Steve. It's a pleasure to be here and discuss the importance of developing critical thinking skills. \\ \hline
James & Hello, Steve. Thank you for having me. I'm looking forward to sharing some practical tips and strategies. \\ \hline
\cellcolor{blue!10}Host & \cellcolor{blue!10}Fantastic! So to kick us off, Dr. Linda, would you mind giving us a brief overview of what critical thinking is and why it's important? \\ \hline
\cellcolor{yellow!10}Dr. Linda & \cellcolor{yellow!10}Certainly, Steve. Critical thinking is the ability to analyze information objectively and make a reasoned judgment. It involves the evaluation of sources, such as data, facts, observable phenomena, and research findings. Good critical thinkers can draw reasonable conclusions from a set of information, and distinguish between useful and less useful details. \\ \hline
\cellcolor{blue!10}Host & \cellcolor{blue!10}That's a clear explanation, thank you, Dr. Linda. How about you, James? As an author, how did developing your critical thinking skills influence your writing process? \\ \hline
\cellcolor{green!10}James & \cellcolor{green!10}Great question, Steve. Critical thinking plays a huge role. Not only does it help in researching and understanding different viewpoints before forming my own, but also in constructing clear, concise, and persuasive arguments. It's like having a good quality control mechanism in your brain! \\ \hline
Host & Quality control for the brain- I like that! We'll continue to ponder this and delve deeper into ways to develop these critical thinking skills after a short break. Stay tuned! \\ \hline
\end{tabular}
\caption{Example of conversation scripts generated by \textbf{directly asking GPT-4} to generate a
conversation script for a provided topic: How can I develop my critical thinking skills?} % Add a caption for the table
\label{tab:directly} % Add a label for referencing the table
\end{table*}

\begin{table*}[htbp]
\centering
\begin{tabular}{|>{\raggedright\arraybackslash}p{3cm}|>{\raggedright\arraybackslash}p{12cm}|}
\hline
\textbf{Speaker} & \textbf{Speaking Content} \\ \hline
Host & It's a pleasure to have you with us on this enlightening journey to uncover the power of the mind. My first guest for today is Dr. Sarah Smith, a renowned psychologist and author, and my second guest is Mr. Peter Green, an innovative education consultant. \\ 
  & \cellcolor{blue!10}Let's start with the basics, Dr. Smith can you enlighten us on what critical thinking is, and why it's important? \\ \hline
\cellcolor{yellow!10}Dr. Sarah Smith & \cellcolor{yellow!10}Certainly. Critical thinking involves objective analysis and evaluation of an issue to form a judgement. It's important as it helps us make informed and rational decisions, and question existing ideas and beliefs. \\ \hline
\cellcolor{blue!10}Host & \cellcolor{blue!10}Very insightful, Dr. Smith. Moving on to you, Mr. Green, could you share some strategies we can use to improve our critical thinking skills? \\ \hline
\cellcolor{green!10}Mr. Peter Green & \cellcolor{green!10}Of course. You can start with questioning everything, look for evidence, understand different perspectives, and practice active listening. And of course, continual learning and practicing these skills will only help you further develop them. \\ \hline
\cellcolor{blue!10}Host & \cellcolor{blue!10}Wonderful inputs, Mr. Green. Dr. Smith, do you have any captivating experiences or examples, to bring the concept of critical thinking to life? \\ \hline
\cellcolor{yellow!10}Dr. Sarah Smith & \cellcolor{yellow!10}Yes, in my experience, I've seen clients who have made drastic changes in their life just by applying critical thinking. One particular client was stuck in a cycle of unhealthy relationships but through the process of critical thinking, she was able to identify the problem areas and make better choices in her relationships. \\ \hline
\cellcolor{blue!10}Host & \cellcolor{blue!10}Impressive transformation indeed! Mr. Green, how does your work help in promoting critical thinking? \\ \hline
\cellcolor{green!10}Mr. Peter Green & \cellcolor{green!10}My education technology startup, ThinkTank, encourages higher-level thinking by providing stimulating content and puzzles. It's all about breaking down complex problems and finding creative solutions using critical thinking. \\ \hline
\cellcolor{blue!10}Host & \cellcolor{blue!10}That's quite commendable, Mr. Green. Lastly, what challenges do we face in teaching or learning critical thinking and how can they be addressed? \\ \hline
\cellcolor{yellow!10}Dr. Sarah Smith & \cellcolor{yellow!10}One major challenge is removing cognitive biases. It's a long process, but being aware of them and deliberately challenging them can help overcome this hurdle. \\ \hline
\cellcolor{green!10}Mr. Peter Green & \cellcolor{green!10}In addition to what Dr. Smith said, there's also a need for environments like schools, workplaces, or even homes to encourage critical thinking instead of accepting information at face value. \\ \hline
\end{tabular}
\caption{Example of Conversation Scripts generated by \textbf{Host-Guest-Writer (Single-Agent)}. Topic: How can I develop my critical thinking skills?} % Add a caption for the table
\label{tab:singleagent} % Add a label for referencing the table
\end{table*}

\begin{table*}[htbp]
\centering
\begin{tabular}{|>{\raggedright\arraybackslash}p{3cm}|>{\raggedright\arraybackslash}p{12cm}|}
\hline
\textbf{Speaker} & \textbf{Speaking Content} \\ \hline
Host & Hello everyone, welcome to our in-depth talkshow 'Think with Us'! I'm your host, Alex, and today we'll be touching the cornerstone of decision-making, the art of Critical Thinking. We have two honored guests joining us tonight—Dr. Jane Martin, a renowned cognitive psychologist, and Prof. Tom Adams, a distinguished philosophy professor. \\ 
  & \cellcolor{blue!10}Welcome to the show Dr. Martin and Prof. Adams! It's great to have you both with us. Dr. Martin, I'll begin with you. Could you briefly explain to our audience what exactly critical thinking is? \\ \hline
\cellcolor{yellow!10}Dr. Jane Martin & \cellcolor{yellow!10}Sure, Alex. Critical thinking is the ability to logically connect ideas, analyze our own thoughts and assess the validity of arguments around us. In an era of the information overload, it helps us make informed, accurate and ethical decisions. \\ \hline
\cellcolor{blue!10}Host & \cellcolor{blue!10}Insightful explanation, Dr. Martin! Now, Prof. Adams, would you suggest that critical thinking can be taught and developed? \\ \hline
\cellcolor{green!10}Prof. Tom Adams & \cellcolor{green!10}Absolutely, Alex. Encouraging curiosity, teaching argument validation, understanding biases, encouraging self-reflection and broadening the horizons of our knowledge can all contribute to the development of critical thinking. All it requires is patience and determination. \\ \hline
\cellcolor{blue!10}Host & \cellcolor{blue!10}Dr. Martin, as Prof. Adams mentioned biases, could you elaborate more on the obstacles that can hinder the development of critical thinking? \\ \hline
\cellcolor{yellow!10}Dr. Jane Martin & \cellcolor{yellow!10}Indeed. Intellectual humility, confirmation bias, cognitive laziness, and navigating the crowded information landscape all pose challenges. Also dealing with the discomfort of conflict between new and old ideas, also known as cognitive dissonance can be an obstacle. \\ \hline
\cellcolor{blue!10}Host & \cellcolor{blue!10}Prof. Adams, would you add some practical tips on improving these abilities? \\ \hline
\cellcolor{green!10}Prof. Tom Adams & \cellcolor{green!10}One should engage with different perspectives, debate with others and always find the counter-argument to their own point of view. Logical puzzles and keeping a journal of decision-making processes can significantly enhance critical thinking. \\ \hline
\cellcolor{blue!10}Host & \cellcolor{blue!10}Those are indeed valuable tips, thank you, Prof. Adams. Dr. Martin, how does your field of cognitive psychology contribute to critical thinking? \\ \hline
\cellcolor{yellow!10}Dr. Jane Martin & \cellcolor{yellow!10}Cognitive psychology provides insights into how we can understand and modify our thinking behaviors. It guides educators on how to teach critical thinking skills and helps individuals understand their own thought processes. We continually strive to improve thinking and enrich lives. \\ \hline
\cellcolor{blue!10}Host & \cellcolor{blue!10}That's fascinating, Dr. Martin! And what would you say, Prof. Adams, is the contribution of your field? \\ \hline
\cellcolor{green!10}Prof. Tom Adams & \cellcolor{green!10}Philosophy is the bedrock of critical thinking. It encourages questioning, exploring diverse perspectives and seeks universal truths. The teachings of great philosophers like Socrates are valuable tools for nurturing critical thinking skills. \\ \hline
Host & That's absolutely enlightening, thank you both for your insights. Ladies and gentlemen, that's all the time we have for today. By embracing the art of critical thinking, we can better equip ourselves to navigate this complex world. Until next time, keep questioning, keep learning. \\ \hline
\end{tabular}
\caption{Example of Conversation Scripts generated by \textbf{Host-Guest-Writer (Multi-Agent)}. Topic: How can I develop my critical thinking skills?} % Add a caption for the table
\label{tab:casestudty_m} % Add a label for referencing the table
\end{table*}

%\section{Human Annotators}
%We recruit human annotators from crowdsourcing platforms (\href{https://www.prolific.com/}{https://www.prolific.com/}.) to conduct subjective tests on voice-role matching and audio comparison, as outlined in Section 4.3. For the voice-role matching test, we engage 6 native speakers for evaluation, while 9 native speakers participate in the speech comparison test. The average cost per participant for each test is approximately \$7 to \$8. 
%The instruction for each test can be found in Figure \ref{fig:match_inst} and Figure \ref{fig:compare_inst}.

%\begin{figure*}[h]
%  \centering
%  \includegraphics[width=\textwidth]{Figures/match.png}
%  \caption{Instruction for Voice-Role Matching Test.}
%  \label{fig:match_inst}
%\end{figure*}

%\begin{figure*}[h]
%  \centering
%  \includegraphics[width=\textwidth]{Figures/compare.png}
%  \caption{Instruction for voice comparison test.}
%  \label{fig:compare_inst}
%\end{figure*}

\end{document}